\documentclass[referee]{aastex61}
\usepackage{graphics,amsmath,amsbsy}
\usepackage{enumitem}

\definecolor{mrk}{RGB}{0,0,0}
\newcommand{\mrk}[1]{{\color{mrk} #1}}

\renewcommand{\vec}[1]{ \protect {{\mathbf{\boldsymbol{#1}}}}}

\shorttitle{Particle acceleration modelling in an individual solar flare}
\shortauthors{Gordovskyy et al.}
\begin{document}

\title{Forward modelling of particle acceleration and transport in an individual solar flare}

\author[0000-0003-2291-4922]{Mykola Gordovskyy}
\affil{Department of Physics \& Astronomy, University of Manchester, Manchester M13 9PL, UK}

\author[0000-0002-7089-5562]{Philippa K. Browning}
\affil{Department of Physics \& Astronomy, University of Manchester, Manchester M13 9PL, UK}

\author[0000-0001-5121-5122]{Satoshi Inoue}
\affil{Institute for Space-Earth Environmental Research, University of Nagoya, Nagoya 464-8601, Japan}

\author[0000-0002-8078-0902]{Eduard P. Kontar}
\affil{School of Physics \& Astronomy, University of Glasgow, Glasgow, G12 8QQ, UK}

\author[0000-0002-6814-6810]{Kanya Kusano}
\affil{Institute for Space-Earth Environmental Research, University of Nagoya, Nagoya 464-8601, Japan}

\author{Grigory E. Vekstein}
\affil{Department of Physics \& Astronomy, University of Manchester, Manchester M13 9PL, UK}

\begin{abstract}
The aim of this study is to generate maps of the hard X-ray emission produced by energetic electrons in a solar flare and compare them with observations. The ultimate goal is to test the viability of the combined MHD/test-particle approach for data-driven modelling of active events in the solar corona and their impact on the heliosphere. Based on an MHD model of X-class solar flare observed on the 8th of September 2017, we calculate trajectories of a large number of electrons and protons using the relativistic guiding-centre approach. Using the obtained particle trajectories, we deduce the spatial and energy distributions of energetic electrons and protons, and calculate bremsstrahlung hard X-ray emission using the 'thin target' approximation. Our approach predicts some key characteristics of energetic particles in the considered flare, including the size and location of the acceleration region, energetic particle trajectories and energy spectra. Most importantly, the hard X-ray bremsstrahlung intensity maps predicted by the model are in a good agreement with those observed by RHESSI. Furthermore, the locations of proton and electron precipitation appear to be close to the sources of helioseismic response detected in this flare. Therefore, the adopted approach can be used for observationally-driven modelling of individual solar flares, including manifestations of energetic particles in the corona, as well as inner heliosphere.
\end{abstract}

\section{Introduction}\label{s-intro}

Observations show that high-energy non-thermal particles may carry a substantial part of the energy released in solar flares \cite[see e.g.][]{asce16,kone19}. Most of the energetic protons and electrons accelerated in flares precipitate in the solar atmosphere, heating it and producing emission in radio, microwave, X-ray and $\gamma$-ray domains. Some of the energetic particles escape from the corona into the heliosphere, and can be observed in-situ, usually at 1~a.u. near the Earth, but now also near the Sun, by Parker Solar Probe and Solar Orbiter missions. Particles accelerated in geo-effective flares and related active events, such as coronal mass ejections, act as an essential link between the Sun and the Earth' and are one of the main ingredients of space weather. Therefore, understanding the physical mechanisms behind particle acceleration and transport in the corona, as well as their escape into the heliosphere are key to understanding the physics of Sun-Earth connections and predicting the space weather. 

Comparison of the energetic electron properties in the corona and near the Earth reveals a strong correlation between their energy spectra and temporal evolution, at least in prompt events \cite[e.g.][]{krue07}, indicating that both energetic electron populations have the same origin. Still, their properties in the corona and at 1~a.u. appear to indicate that both populations are strongly affected during transport through the corona and the heliosphere. Until now, it was almost impossible to distinguish between different effects affecting particle transport in the corona and in the heliosphere, because reliable observational diagnostics of electrons and ions were possible only in the lower solar atmosphere, through the radio-, X-ray and $\gamma$-ray emission they produce, or near the Earth, using in-situ observations. However, in the next few years Parker Solar Probe (PSP) and Solar Orbiter (SolO) missions will provide remote (low-frequency radio) and in-situ observations of solar energetic particles much closer to the Sun, in the inner heliosphere. In order to effectively use the capabilities of these instruments and exploit the data, it is necessary to model magnetic reconnection and energy release, particle acceleration and transport, both towards the chromosphere and into the heliosphere, in individual solar flares.

Large-scale modelling of active events in the solar atmosphere is usually performed using the MHD approach, which does not account for non-thermal plasma. However, kinetic approaches, which can consistently account for non-thermal particles, cannot be realistically used at large scales. Hence, an approach combining MHD and kinetic simulations is often used used \citep[see e.g.][for review and references]{gore19}. Hybrid fluid-kinetic methods can provide more rigorous and self-consistent description of energetic particles at large scales. However, they are computationally-expensive, and cannot be practically used for modelling all particle species in a configuration with the unknown topology of the 'kinetic' regions, i.e. regions with high numbers of energetic particles, that need to be treated using a kinetic approach. Therefore, combination of MHD and test-particle methods (MHDTP approach, thereafter) is, perhaps, the optimal way of modelling individual solar flares at large scales. 

The MHDTP approach has been extensively used to investigate energetic particle kinetics in various generic magnetic reconnection models, starting from the simulations of particle motion in quasi-stationary 2D models \cite[e.g.][]{saka90, klie94, vebr97,brve01}. For instance, \citet{zhgo04} and \citet{wone05} used quasi-stationary 2D field configurations to reveal the possibility of electron-ion separation in reconnecting current layers with strong guiding field. Three-dimensional models based on quasi-stationary electromagnetic fields were used to investigate particle motion in the vicinity of 3D null-points \cite[e.g.][]{dabr05,dabr06,guoe10,pont11} revealing fundamental differences from particle acceleration in nearly two-dimentional current layers. Later, more complicated, time-dependent MHDTP models have been used to study particle acceleration and transport in more realistic 2D and 3D configurations of reconnecting current layers in the solar corona and Earth's magnetosphere \cite[e.g.][]{bire04,gore10a,gobr11,zhoe15,zhoe16}. For instance, \citet{ture05} and \citet{gobr11,gobr12} used 3D MHDTP simulations to show that diffuse, large-scale acceleration regions with fragmented, stochastic electric field configurations can be efficient particle accelerators, and an alternative to the standard solar flare scenario \citep{shie95}. \citet{gore14,gore16,gore17} and \citet{pine16} successfully used MHDTP modelling to investigate solar flares involving unstable twisted coronal loops, and compare the evolution of thermal and non-thermal emissions associated with these events.

MHDTP simulations involving observationally-driven MHD models are a substantial step forward. A number of studies has developed observationally-driven MHD models of solar active events in the last two decades. \citet{guno05a,guno05b} considered an MHD model of an active region using the potential coronal field reconstruction as an initial condition, and produced observables -- synthetic thermal EUV emission -- which could be directly compared with observations. Later, \citet{boue13} used the same strategy to model magnetic field and plasma evolution in the corona above a non-flaring active region. The predicted EUV intensity maps and EUV Doppler line shifts were found to be in a relatively good agreement with observations. These MHD simulations were later used by \citet{thre16} to model particle acceleration. They have demonstrated how particles are accelerated in that configuration, although in their model a substantial uniform resistivity resulted in high electric field occupying a substantial part of the domain, which, in turn, yielded a very high particle acceleration efficiency.

However, the magnetic field structure in an active region can be very different from the potential field, particularly before active events (flares or eruptions), that is why initialising MHD models of actual events with non-linear force-free (NLFF) field extrapolation is a significant step forward. Using the NLFF extrapolation method developed by \citet{inoe11}, a series of observationally-driven MHD models have been constructed to investigate triggering of solar flares \citep[e.g.][]{muhe17} and onset of eruptions in actual active regions \citep[e.g.][]{inoe18a,wooe18}.

There are two different approaches to the observationally-driven modelling. While the majority of studies use the coronal magnetic field reconstructions only to initialise the event, some studies use ongoing driving, using the evolving photospheric magnetic field as an active boundary condition \cite[e.g.][]{jiae16}. The latter approach is valid for modelling slowly-evolving corona at longer temporal scales, but is difficult to use for modelling fast transient events, such as flares. The main reasons are that in the fast evolving corona the magnetic field may be substantially different from force-free, and the magnetic field observations at the photosphere can be unreliable due to flare-related changes in spectral lines used for measurements.

In this study we combine observationally-driven MHD simulations of an individual solar flare \citep{inoe18b} with test-particle simulations, using the method developed by \citet{gore10b,gore14} in order to investigate proton and electron acceleration and transport. The aim is to produce synthetic observables associated with the energetic particles (particularly, electrons), which can be directly compared with existing and forthcoming observations. The ultimate goal is to test the approach by comparing synthetic observables with the actual data for this and other events. The observations and methodology are described in Section~\ref{s-model}, the results are shown in Section~\ref{s-results} and summarised in Section~\ref{s-summary}.

\begin{figure*}[ht]
\centerline{\includegraphics[width=0.9\textwidth,clip=]{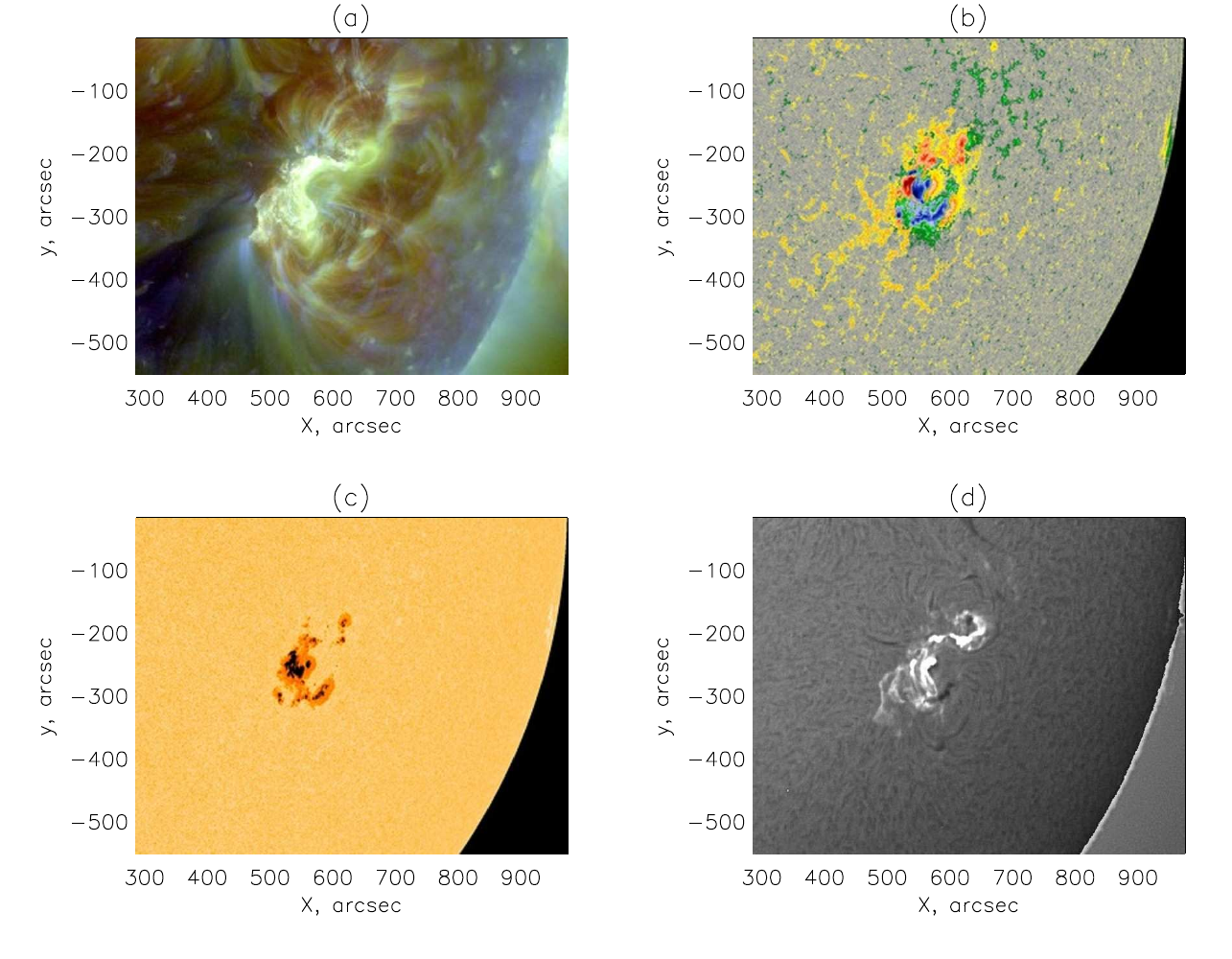}}
\caption{The active region NOAA AR 12673, where the 6th of September 2017 flare occurred. Panel (a) shows combined SDO/AIA intensity map before the flare (at 11:40~UT) including emission in 211$\mathrm{\AA}$ (red), 193$\mathrm{\AA}$ (green) and 171$\mathrm{\AA}$ (blue) lines. Panel (b) shows SDO/HMI magnetogram before the flare (at 11:30~UT). Panel (c) shows SDO/AIA continuum intensity before the flare (at 11:30~UT). Panel (d) shows NSO Cerro Tollo GONG H$\alpha$ intensity map at the beginning of the thermal phase (at 12:25~UT).}
\label{f-ar1}
\end{figure*}

\begin{figure*}[ht]
\centerline{\includegraphics[width=0.75\textwidth,clip=]{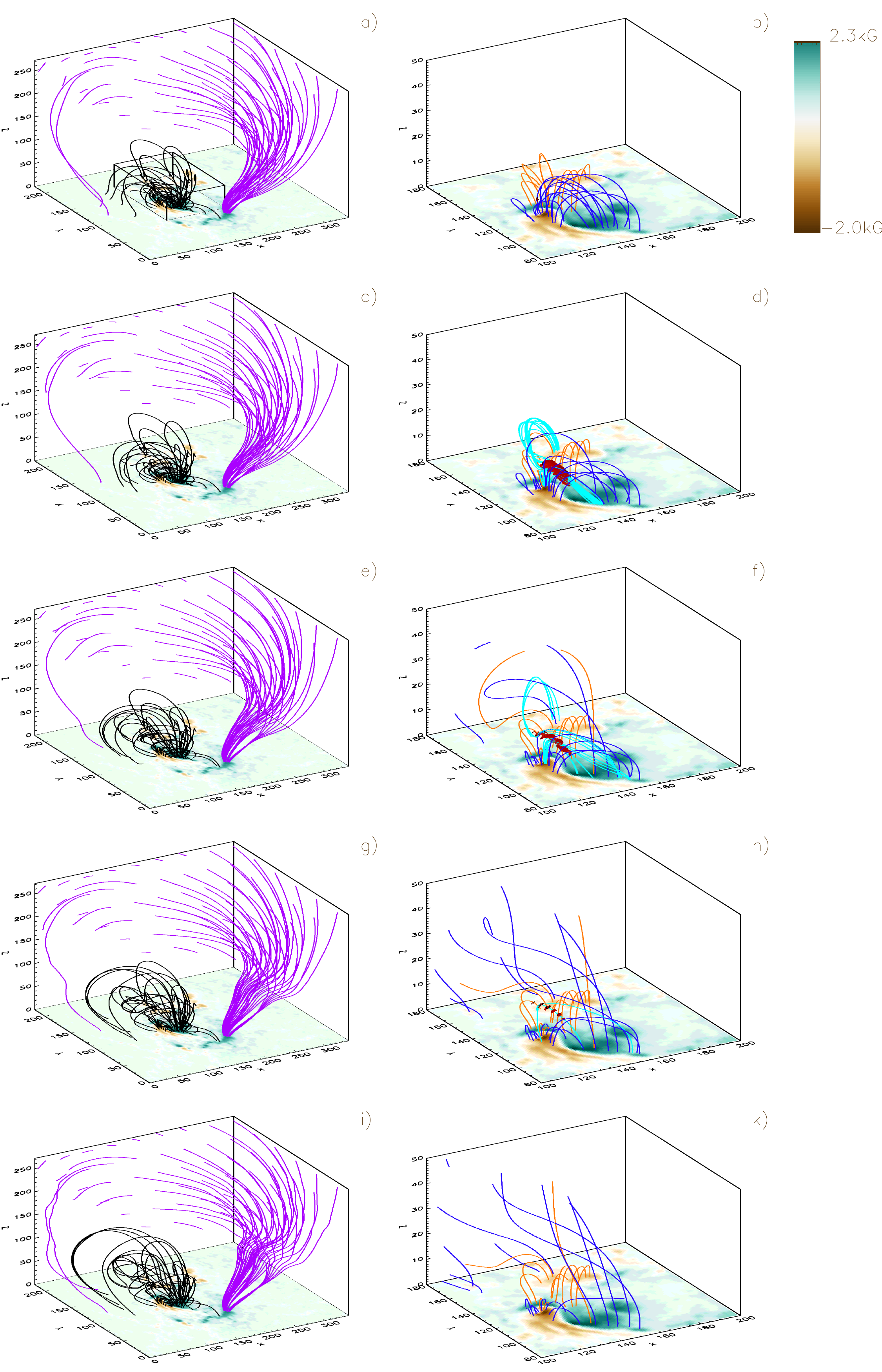}}
\caption{Evolution of the magnetic field in the considered active region. \mrk{Left panels show the whole domain (224.8$\times$158.4$\times$195.8~Mm$^3$), while right panels show smaller volume of 72$\times$72$\times$36~Mm$^3$ denoted by a black box in panel (a). Left panels show selected field lines originating from the strong bipolar region containing two sunspots (black lines) and those, origination from the upper boundary of the domain (purple lines)}. Right panels show a subset of the model with the diffusion region defined as per Equation~10 (brown surface), \mrk{ field lines going through the diffusion region (light blue lines), field lines connection the area of positive polarity with NW and SE areas of negative polarity (orange and dark blue lines, respectively)}. Panels (a-b), (c-d), (e-f), (g-h) and (i-k) correspond to times 0~s, 80~s, 249~s, 415~s, and 581~s, respectively. \mrk{The tick labels are in units of 0.72~Mm.} }
\label{f-mhd}
\end{figure*}

\section{Observationally-driven modelling}\label{s-model}

The approach used here is based on consecutive application of three different methods: (1) reconstruction of the coronal magnetic field assuming it is force-free using the photospheric magnetic field observations, (2) MHD simulations of electromagnetic field and thermal plasma evolution, using the reconstructed coronal magnetic field as an initial condition, and (3) calculation of large number of test-particle trajectories in evolving electric and magnetic field obtained in MHD simulations. The latter method provides information about the evolution of electron and proton distributions in the phase space and is used for calculation of synthetic observables. Here we consider hard X-ray bremsstrahlung emission produced by energetic electrons bombarding dense chromosphere and photosphere, although observables in other spectral domains, such as microwave, can be also derived \citep{shku16,gore17}.

\subsection{The 6th of September 2015 flare observations}\label{s-obs}

This flare was observed in the active region NOAA 12673, which produced a series of powerful solar flares over several days (Figure~\ref{f-ar1}). The flare occurred on the 6th of September 2017, when the active region was in the SW quarter of the solar disk, at the projection angle (the angle between the line-of-sight and the normal to the solar surface) of around 36$^{\circ}$. It started at 11:52~UT, peaking (in X-ray) at 12:02~UT, with the thermal phase (in H$\alpha$) lasting until approximately 15:30~UT.

The active region had a complex structure with mixed magnetic polarities (Figure~\ref{f-ar1}b). A large group of sunspots in a form of two clusters with opposite polarities was located in the NE part of the active region (Figure~\ref{f-ar1}c). These sunspots had a joint penumbra and magnetic fields of 1.5--2~kG. The sunspots of opposite polarities were separated by an S-shaped neutral line. Interestingly, sunspot magnetic field measurements at Mt.Wilson 15-foot telescopes approximately 36 hours later show that this sunspot cluster became unipolar, which could be the result of magnetic field topology change during a series of flares, which occurred on the 6th and 7th of September.

\begin{figure}[ht]
\centerline{\includegraphics[width=0.25\textwidth,clip=]{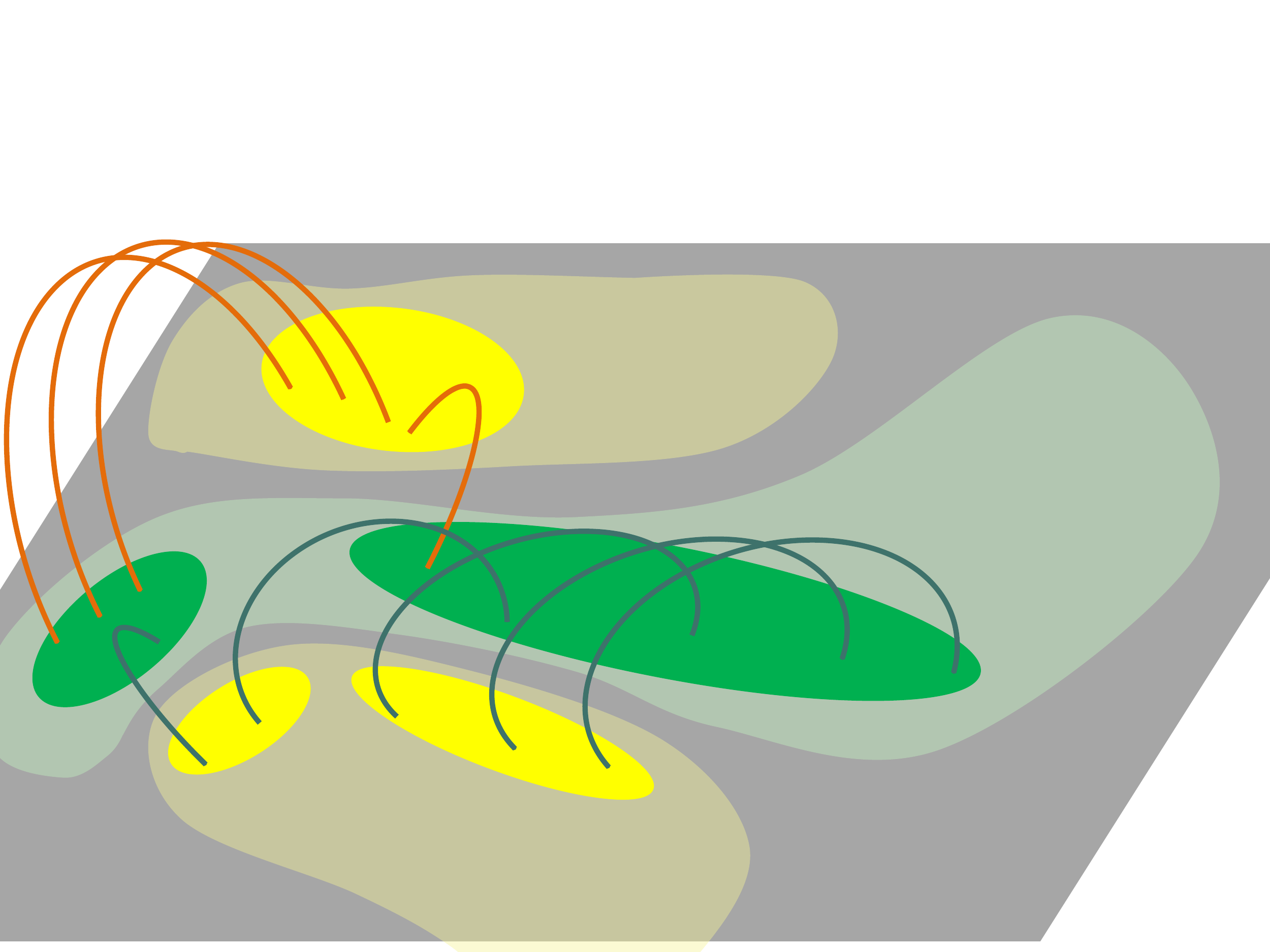}}
\centerline{\includegraphics[width=0.25\textwidth,clip=]{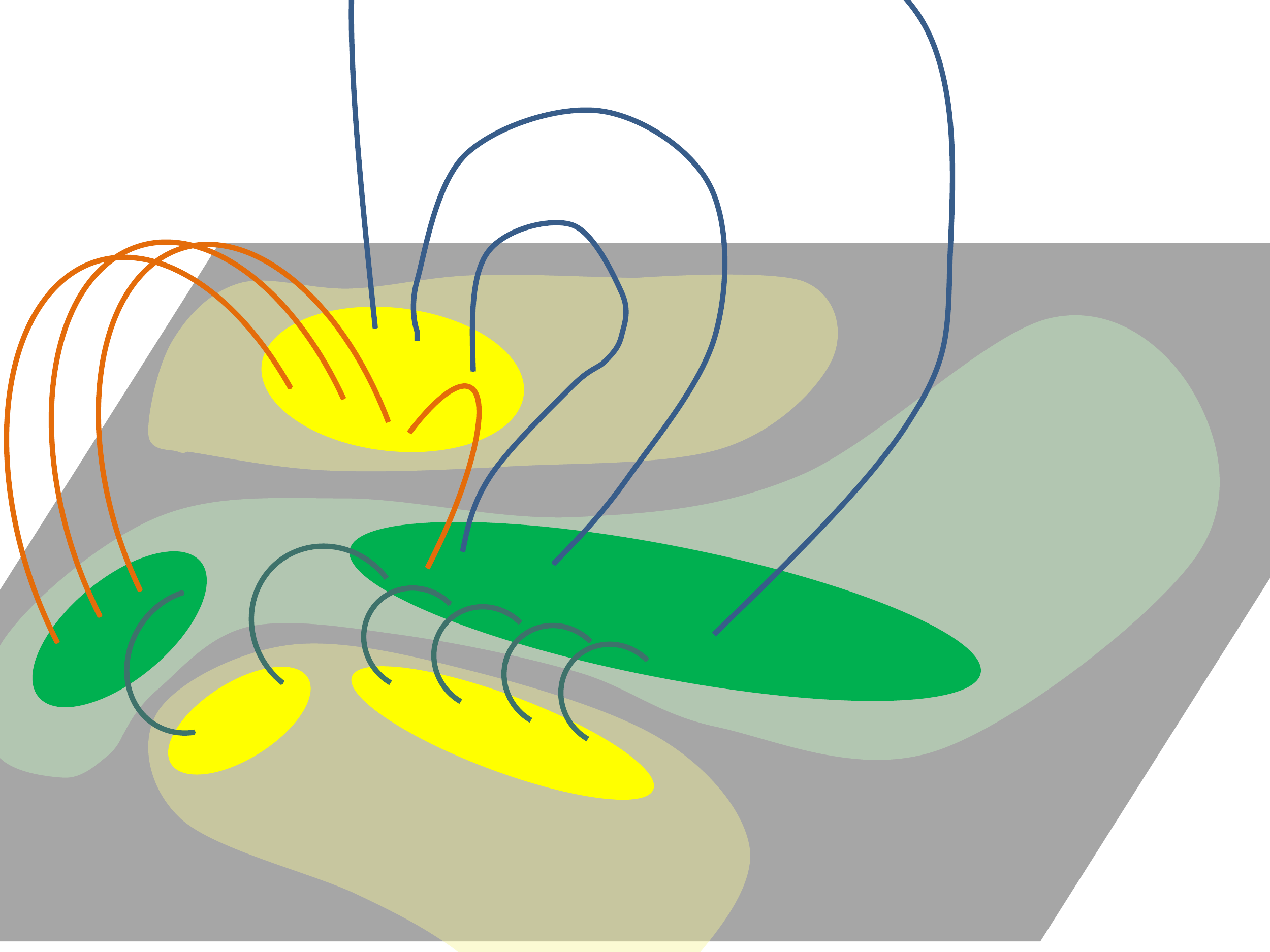}}
\caption{\mrk{Cartoons showing the magnetic connectivity before (top) and after (bottom) the event. The green area corresponds to the strong positive magnetic polarity area in right panels in Figure~2, while yellow areas represent NW and SE areas of negative polarity in right panels in Figure~2.}}
\label{f-sketch}
\end{figure}

Two smaller sunspots of opposite polarities were located in the south-eastern part of the active regions. Furthermore, the active region included two diffuse areas with clusters of small sunspots, one with positive magnetic field polarity at the north, another with negative polarity at the south-west.

At the beginning of the thermal phase (around 12:30~UT), H$\alpha$ intensity maps revealed two bright elements: two ribbons on the each side of the neutral line separating the large sunspots, and a single arc-shaped ribbon connecting the two large sunspots with the area of positive polarity in the northern part of the active region (Figure~\ref{f-ar1} d).

\subsection{Magnetic field reconstruction and MHD simulations of solar flares}\label{s-mhd}

The magnetohydrodynamic part of this model is based on a study by \citet{inoe18b}. The observed event is located well away from the limb and the relatively small projection angle for this area means that the vector magnetograms are of sufficient quality for non-linear force-free (NLFF) coronal magnetic field reconstruction. Hence, the initial magnetic field configuration in the considered model has been derived by solving $\left ( \vec{\nabla} \times \vec{B} \right ) \times \vec{B} = 0$ equation numerically, based on the photospheric magnetic field taken at 08:36~UT, well before the flaring even, approximately 20~minutes prior to a previous, smaller flare in this group.

\mrk{Since the initial magnetic field configuration is non-potential, it starts evolving, relaxing towards the configuration with lower magnetic energy, when the resistivity is switched on.} Evolution of the magnetic field and thermal plasma in this model is considered using the following set of MHD equations for magnetic field $\vec{B}$, velocity $\vec{v}$ and current density $\vec{j}$ as follows:

\begin{eqnarray}
\frac {\partial \vec{v}}{\partial t} &=& - \left ( \vec{v} \cdot \vec{\nabla} \right ) \vec{v} + \frac 1\rho \vec{j} \times \vec{B} + \nu \nabla^2 \vec{v} \label{eq-mhdv}\\
\frac {\partial \vec{B}}{\partial t} &=&  \vec{\nabla} \times \left ( \vec{v} \times \vec{B} - \eta \vec{j} \right ) - \vec{\nabla} \phi \label{eq-mhdb}\\
\vec{j} &=& \vec{\nabla} \times \vec{B}, \label{eq-mhdj}
\end{eqnarray}
where $\vec{v}$, $\vec{B}$ and $\vec{j}$ are the velocity, magnetic field and current density, respectively. (All equations and values in Section~2.2 are dimensionless.) Constant uniform dissipation coefficients are used in this MHD model, resistivity $\eta=1.0\times10^{-5}$ and viscosity $\nu=1.0\times10^{-4}$. The last term in Equation~\ref{eq-mhdb} is used in order to reduce spurious numerical $\vec{\nabla} \cdot \vec{B}$. The potential $\phi$ is calculated using the following equation \citep{dede02}:

\begin{equation}
\frac {\partial \phi}{\partial t} = - c_h^2 \vec{\nabla} \times \vec{B} - \frac {c_h^2}{c_p^2} \phi \label{eq-mhdp},
\end{equation}
with constants $c_h^2=0.04$ and $c_p^2 =0.1$.

The main simplification of this approach, compared with the standard one-fluid MHD, is that it assumes that plasma beta is negligibly low, and, hence, gas pressure can be ignored. \mrk{This is a reasonable assumption, as in our model $\beta$ is expected to be around $10^{-3}$ in the quiet corona ($T\approx10^6$~K), rising to $\sim10^{-2}$ in and around the energy release region ($T\approx10^7$~K).} Furthermore, in this model, the dimensionless thermal plasma density $\rho$ is assumed to be proportional to the dimensionless local magnetic field strength throughout $\rho = |\vec{B}|$. \mrk{Although, qualitatively, this reflects the structure of the plasma density in the corona, this assumption means that we cannot reliably study the dynamics of the thermal plasma close to the energy release region. However, since the plasma $\beta$ is negligibly low, this assumption does not affect the evolution of electromagnetic fields in our model and, hence, does not affect the particle simulations using the MHD model.}

The normal component of magnetic field is fixed on all six boundaries of the computational domain, \mrk{while the tangential components are calculated using the induction equation~(\ref{eq-mhdb})}. All components of velocity are set to zero on all boundaries. Neumann boundary conditions are used for function $\phi$: $d\phi/d\vec{n} = \vec{0}$.

\mrk{The above set of the MHD equations is solved on a uniform Cartesian grid of 340$\times$220$\times$272 elements. It has dimensionless size of 1~L$_0$~$\times$~0.647~L$_0$~$\times$~0.8~L$_0$ or 224.8~Mm~$\times$~158.4~Mm~$\times$~195.8~Mm in real units, and the grid step is 2.9$\times$10$^{-3}$~L$_0$ or 0.72~Mm in all three dimensions. The scaling values for the magnetic field and time are $B_0=2.5$~kG and $t_0=147$~s, respectively. Hence, the uniform resistivity used in the MHD simulations corresponds to the diffusion coefficient of approximately $4\times10^9$~m$^2$~s$^{-1}$ or to the Lundquist number of about 10$^5$.}

\subsection{Particle simulations and non-thermal emission}\label{s-part}

For typical magnetic field values in the solar corona ($1-100$~G) gyro-periods for electrons and protons are order of $10^{-8}-10^{-6}$~s and $10^{-5}-10^{-3}$~s, respectively. The Larmor radii for thermal and accelerated (to $\sim$1~MeV) electrons are $10^{-3}-10^{-1}$~m and $0.1-10$~m, respectively, while Larmor radii for thermal and accelerated ions are $1-100$~m and $10^{2}-10^{4}$~m, respectively. \mrk{These values are much smaller than corresponding spatial grid step in the MHD model (order of $10^6$~m) and timescale of fast magnetic field variations (order of 10~s). Therefore, particle trajectories can be approximated by the trajectories of their gyro-centres and the guiding centre approximation can be used.}

We use the following set of equations to describe guiding centre motion \citep{nort63,gore10a,gobr11}:

\begin{eqnarray}
\frac {d \vec{r}}{dt} &=& {\vec{u}} + \frac{(\gamma V_{||})}\gamma {\vec{b}}\\
{\bf U} &=& {\bf U}_E + \frac mq \frac {(\gamma V_{||})^2}{\gamma \kappa^2 B} [{\bf b} \times ({\bf b}\cdot{\bf \nabla}){\bf b}] + \nonumber \\%
&\,& \, \frac mq \frac \mu {\gamma \kappa^2 B} [{\bf b} \times ({\bf \nabla} (\kappa B))]\\
\frac{d (\gamma V_{||})}{dt} &=& \frac qm {\bf E}\cdot {\bf b} - \frac \mu \gamma ({\bf b} \cdot {\bf \nabla}(\kappa B))+ \nonumber \\
&\,& \, (\gamma V_{||}) {\bf U}_E\cdot (({\bf b}\cdot{\bf \nabla}){\bf b})\\
\gamma &=& \sqrt{\frac{c^2 +(\gamma V_{||})^2 + 2 \mu B}{c^2 - U^2}}\\
\frac{d\mu }{dt} &=& 0.
\end{eqnarray}

Here $\vec{r}$, $V_{||}$ and $\vec{U}$ are guiding centre position, parallel (to the magnetic field) velocity and drift velocity, respectively, $\mu= V^2_{gyro}/(2 B)$ is the specific magnetic moment of a particle, $q$ and $m$ are the charge and mass, respectively, $\vec{b}=\vec{B}/B$ is the magnetic field direction vector. $\vec{U}_E$ is the so-called E$\times$B drift, $\vec{U}_E= [\vec{E} \times \vec{B}]/B^2$. The relativistic parameters are defined as $\kappa = \sqrt{1-U_E^2/c^2}$ and $\gamma = 1/ \sqrt{ 1-V^2/c^2 }$. The full particle velocity can be calculated as $V = \sqrt{V_{||}^2 + |\vec{U}|^2 + (2 \mu B)^2}$, while the particle kinetic energy can be calculated as $\varepsilon = (\gamma-1) mc^2 \approx \frac 12 m V^2$.

\begin{figure*}[ht]
\centerline{\includegraphics[width=0.65\textwidth,clip=]{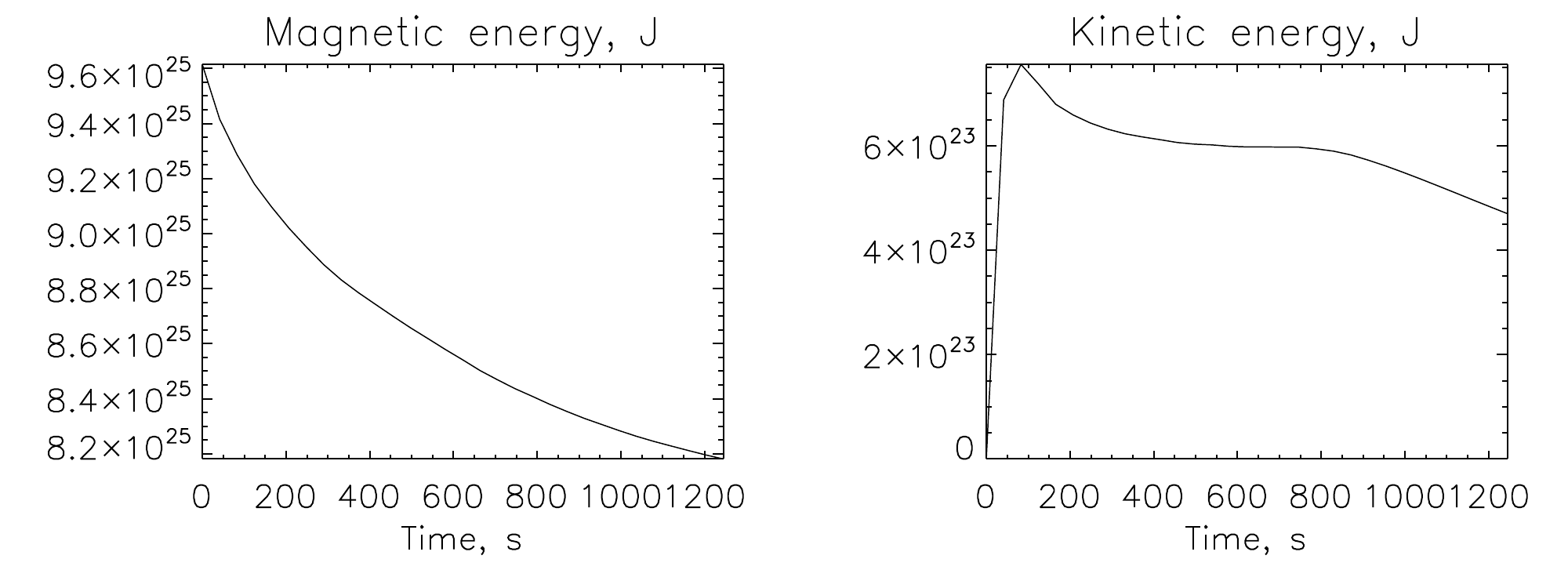}}
\caption{Total magnetic and kinetic energy variation.}
\label{f-nrg}
\end{figure*}

Although we include a number of $d\vec{v_{||}}/dt$ and drift terms in our motion equations, \mrk{the most important terms in Equations 6 and 7 are the $\vec{U}_E$ drift (representing bulk thermal plasma motion) and $E_{||}$ acceleration in a region where anomalous resistivity is sufficiently high}. Other terms are very small, due to relatively weak curvature of the field. 

The magnetic field is taken directly from the MHD simulations. The electric field is calculated using the Ohms law
\[
\vec{E} = - \vec{v} \times \vec{B} + \eta \vec{j}.
\]
The resistivity used in MHD simulations is not appropriate for particle simulations because it includes high background resistivity required to regularise the solution and, hence, would result in very substantial overestimation of the particle energy gain \cite[see e.g.][]{thre16}. This difference between the electric resistivity in MHD and particle simulations is inevitable, because the  spatial resolution of a large-scale MHD model ($10^5-10^6$~m) is, obviously, insufficient to resolve real spatial scales of current density variation in the solar corona $10-10^3$~m \cite[see discussion in][]{gore14}. In test-particle simulations we use anomalous resistivity $\eta$ as defined in \citep{gobr11,gore14}:

\begin{equation}
\eta= 
\begin{cases}
    \eta_0+\eta_1,& |j| > j_{\mathrm{cr}}(\rho, T)\\
    \eta_0,              & \text{otherwise}
\end{cases}.
\end{equation} 

\mrk{The functional form of $j_{\mathrm{cr}}(\rho, T)$ is such that it mimics the anomalous resistivity due to ion-acoustic turbulence and takes into account the resolution of the numerical grid. In the present model, $\eta_1=10^{-3}$ (using the same scaling as in the MHD model), in other words, the corresponding Lundquist number is 10$^3$, and $\eta_0=0$. Taking into account that the temperature of thermal plasma is ignored in the MHD simulations, we assume it equal 1~MK throughout the domain, the critical current threshold is set to $j_{cr}=0.18 \rho$ (in dimensionless units).}

\begin{figure*}[ht]
\centerline{\includegraphics[width=0.99\textwidth,clip=]{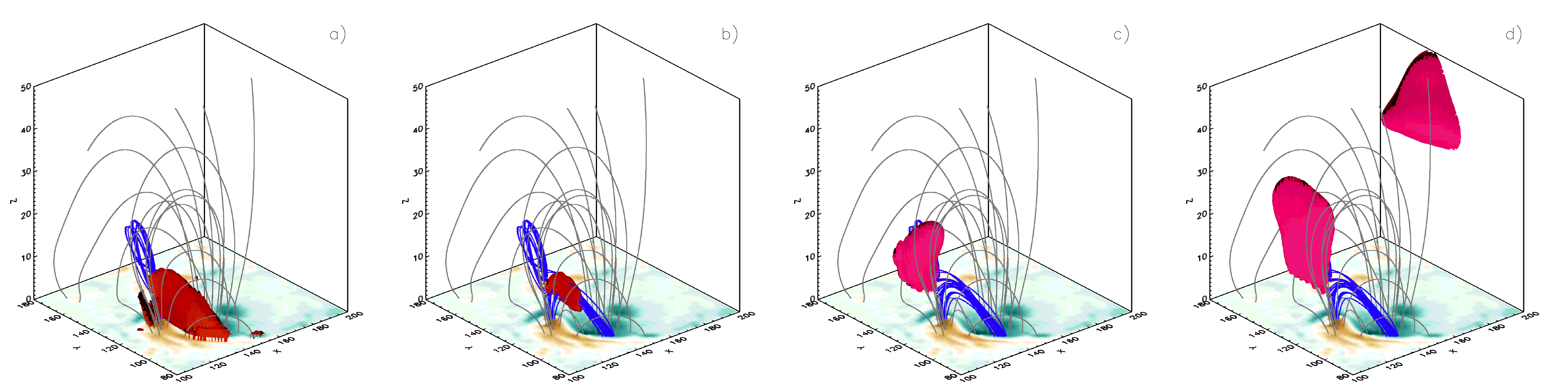}}
\caption{Panel (a): Location of the layer with high parallel current density. Panel (b): locations of the region with high current-driven resistivity, defined by Equation~(10) (acceleration region in test-particle simulations). Panels (c) and (d) show regions with high parallel (to $\vec{B}$) and perpendicular velocities in MHD simulations, respectively. The iso-surfaces in panels (a), (c) and (d) correspond to half-maximum values. All panels correspond to time moment $t=80$~s. \mrk{The tick labels are in units of 0.72~Mm.} }
\label{f-flow}
\end{figure*}

In the particle simulations we trace up to 2$\times$10$^6$ test-particles for each species. The initial spatial distributions of particles are uniform, filling the whole domain. The energy distribution in Maxwellian with a temperature of 0.9~MK throughout the domain. We set 'thermal bath' conditions on all six boundaries: for each particle leaving the domain, another particle of the same species with the velocity randomly chosen from Maxwellian distribution is injected at the same location. This set of initial and boundary conditions is optimal, making the test-particle populations statistically-representative of their species.

The bremsstrahlung HXR emission produced by energetic electrons is calculated using the thin-target approximation: it is assumed that energy distribution of electrons producing HXR in the dense plasma is the same as the distribution of electrons hitting the lower boundary. We use Kramers simplified formula to calculate HXR intensities \cite[see e.g.][]{broe03}:

\begin{equation}
I(\epsilon_{ph}) = \mathrm{const} \int \limits_{\epsilon_{ph}}^{\infty} N(\varepsilon)\frac 1{\epsilon_{ph} \varepsilon} d\varepsilon,
\end{equation}
where $\epsilon_{ph}$ is the photon energy, $\varepsilon$ is electron energy, and $N(\varepsilon)$ is the energy distribution of electrons hitting the lower boundary. The constant in front of the integral, \mrk{which is a combination of fundamental constants and geometric parameters, is taken to be unity}. However, because we are not looking at absolute values of HXR flux in this study, it is taken to be unity, and the resulting spatial and energy distribution of HXR emission in Section~\ref{s-results} are given in arbitrary units.

\section{Results and discussion}\label{s-results}

The magnetic field evolution in this event has been studied in detail by \citet{inoe18b}. The magnetic field in the considered active region is shown in Figure~\ref{f-mhd}. It can be seen that the main changes occur above the neutral line, separating two large sunspots of the opposite polarities. The magnetic field above this part of the active region is produced primarily by three areas: an area of positive polarity containing one large sunspot, an area of negative polarity in the south-eastern part of the region, also containing a large sunspot, and a diffuse area of negative polarity in the north-west part. Hence, we can identify two distinctive structures: magnetic flux connecting the area of positive polarity with the NW negative area and magnetic flux connecting the area of positive polarity with the negative area at SE (orange and grey lines, respectively). The latter undergoes a major change during this event. 

The change of connectivity is shown schematically in Figure~\ref{f-sketch} \cite[see also Figure~3 in][]{inoe18b}. Before the onset of reconnection, the field connecting the positive magnetic area with the SE negative area has a form of a strongly sheared arcade over the neutral line separating the two areas. During the event, this magnetic field transforms into a less sheared, lower arcade with height of around 5~Mm. In addition, part of this magnetic flux switches from the SE negative area to NW negative area, while also undergoing an eruption. Thus, the magnetic field connecting the area of positive polarity with the negative area in NW corner rises from few tens of megametres to about 100~Mm within approximately 600~s, {\it i.e.} its velocity is around 0.08 times the Alfv\'{e}n velocity ($v_A \approx 1.7\times 10^6$~m~s$^{-1}$).

\mrk{The fast energy release occurs during the first 200--300~s after the start of reconnection. This is in a good agreement with the typical duration of an impulsive phase in a large flare. Taking into account that the reconnection timescale in the MHD model depends primarily on the magnetic diffusivity value, one can conclude that the chosen resistivity value of $10^{-5}$ is adequate for our model.}

The uniform resistivity in the MHD simulations results in bulk dissipation of the magnetic energy and gradual acceleration of thermal plasma (Figure~\ref{f-nrg}), producing large-scale flows which gradually subside with time. Approximately, 1.5$\times$10$^{25}$~J of magnetic energy is released during this event, which is in a good agreement with the total energies of X-class flares. Nearly 5\% of the released energy is converted into the kinetic energy.  

Although the magnetic field undergoes changes nearly everywhere in the model domain, the magnetic reconnection and primary energy release occur in a relatively small volume just above the two large sunspots. Our earlier simulations showed that on these large spatial scales, particles are predominantly accelerated by the parallel electric in the diffusion regions. The location and structure of the reconnection region and flows around it is shown in Figure~\ref{f-flow}. It can be seen that the parallel component of the current density forms a relatively thin, vertical layer under the arcade connecting the positive area with SE negative area. The volume with high resistivity as per Equation (10) (which is used in particle simulations), is slightly wider and is also limited vertically. This region is located 14--17~Mm above the photosphere; it extends for approximately 25~Mm southward from the X-point and has vertical and E-W dimensions of approximately 5~Mm. 

The volume around the reconnection region also features some  fast localised plasma flows with the speed of up to 5$\times$10$^4$~m~s$^{-1}$. Notably, there is an outflow from the reconnection region, blowing upwards in the NE direction.

\begin{figure*}[ht]
\centerline{\includegraphics[width=0.85\textwidth,clip=]{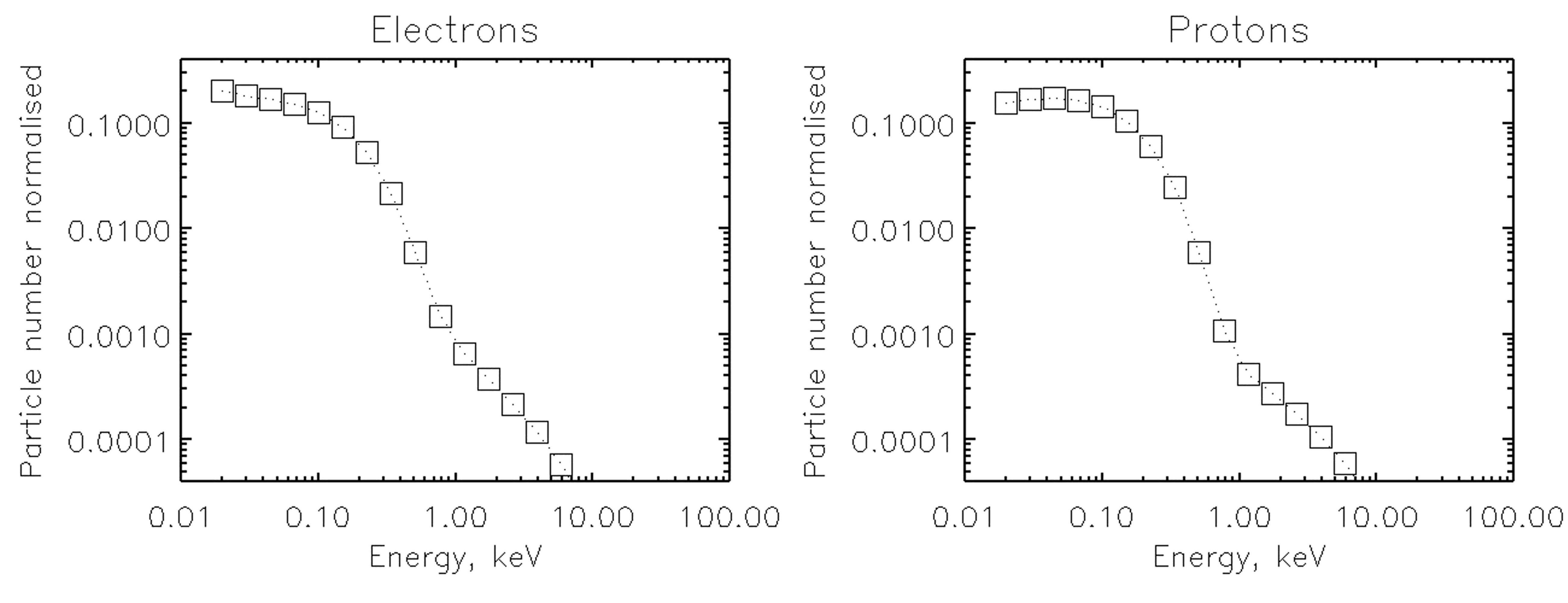}}
\caption{Energy spectra of energetic particles at $t=80$~s.}
\label{f-spectra}
\end{figure*}

The region of fast energy release has a volume which is approximately 4$\times$10$^{-5}$ of the total domain volume. The majority of test-particles never pass through it and, hence, only a small fraction of particles get accelerated. Thus, the initial distribution is made of approximately 2$\times$10$^6$ test-particles for each species. During the simulations, approximately 3.8$\times$10$^{3}$ of test-electrons get above 2~keV, and 7.4$\times$10$^{2}$ test-electrons get energies higher than 10~keV. Most accelerated particles leave the domain within the first 100--120~s. 

The energy spectra of electrons and protons are shown in Figure~\ref{f-spectra}. The spectra are combinations of the Maxwellian cores and a nearly power-law tails at higher energies. Particle energy spectra are very hard, with spectral indices of around 1.9 for electrons and 1.4 for protons. This is likely to be model related: most test-particle models produce spectra with spectral indices in the range of 1--3.

Knowing the corresponding physical volume of the domain (8$\times$10$^{24}$~m$^{3}$), we can estimate the real number of energetic particles, assuming that the average initial plasma density is $3\times10^{-12}$~kg~m$^{-3}$ (approximately $2\times10^{9}$~cm$^{-3}$). Thus, in our model around 6$\times$10$^{36}$ electrons are accelerated above the energy of 10keV. The energy carried by electrons above 10~keV is approximately 1.2$\times$10$^{23}$~J. This is relatively low for an X-class flare, which are normally supposed to produce around 10$^{37}$--10$^{38}$ energetic electrons carrying more than $10^{24}$~J \citep{asce16,kone19}. However, \mrk{the acceleration efficiency in the model strongly depends on the size of the acceleration region and the electric field strength, which, in turn, depend on the parameters defining the functional form of the anomalous resistivity} (Equation 10), such as the threshold current density and the anomalous resistivity amplitude. 

\begin{figure*}[ht]
\centerline{\includegraphics[width=0.55\textwidth,clip=]{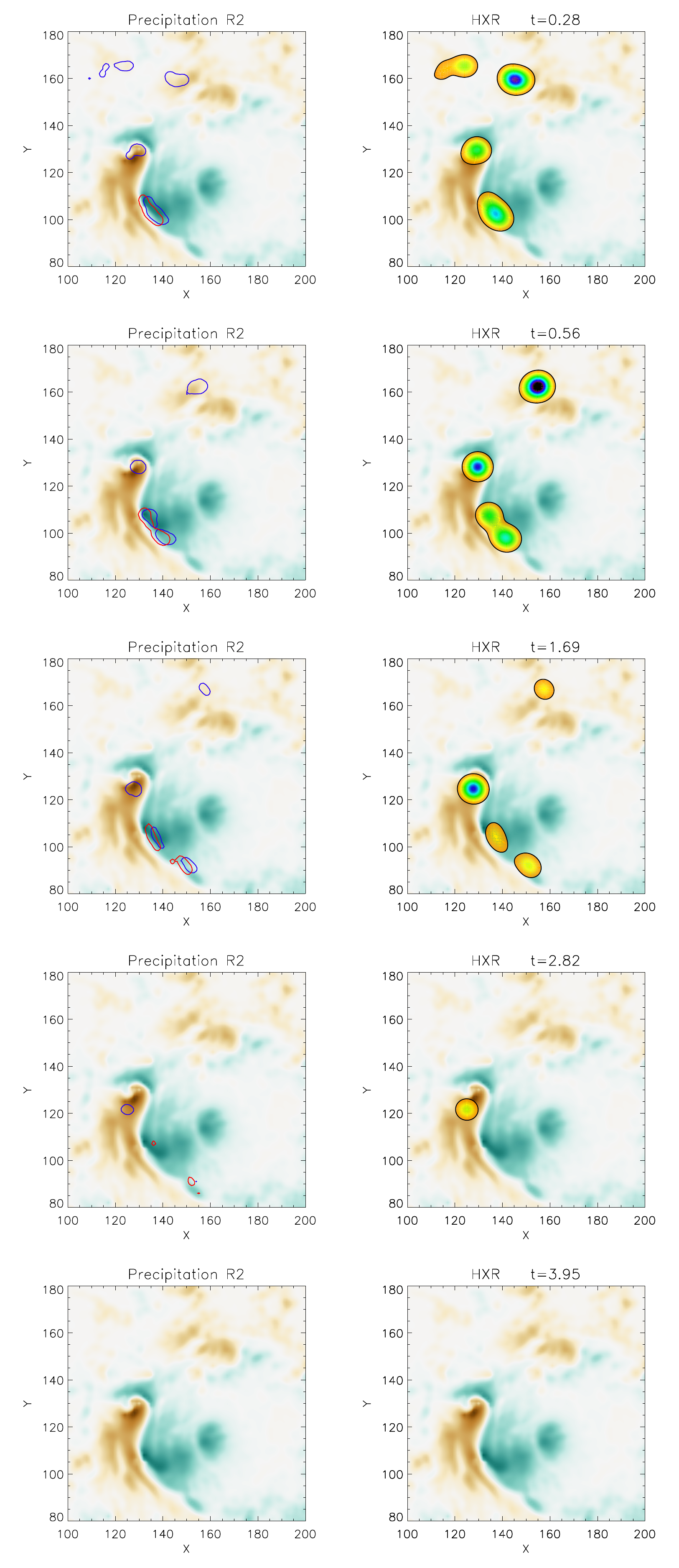}}
\caption{Precipitation sites of energetic particles and the bremsstrahlung emission produced by energetic electrons at different times. Left panels show locations where the majority (96\%) of energetic electrons (blue contours) and protons (red contours) precipitate. Right panels show the intensity maps of the bremsstrahlung hard X-ray emission at 8~keV. The intensity maps have the same scaling and the lower cut-off value corresponds to 20\% of the maximum. The panels, from top to bottom correspond to 0~s, 80~s, 249~s, 415~s, and 581~s, respectively. \mrk{The tick labels are in units of 0.72~Mm.}}
\label{f-particles}
\end{figure*}

The fraction of energetic particles, managing to leave the domain through the upper boundary is negligible, less than 1\%. This is because the magnetic field involved in reconnection is closed at the photosphere. Those few particles which manage to escape upwards are likely to drift from the reconnection region in the south-west direction and, eventually, get into the open field (red magnetic field lines in Figure~\ref{f-mhd}). The large majority of particles precipitate to the photosphere (Figure~\ref{f-particles}). Since there is no particle-particle or kinetic wave scattering, electrons and protons precipitate predominantly to the opposite field polarities. Thus, protons precipitate (i.e. cross the lower, chromospheric boundary of the domain) to the positive polarity, most of them reach the chromosphere near the large positive sunspot. At the same time, electrons precipitate predominantly to the negative magnetic field areas: the sunspot of the negative polarity within the negative SE area and the diffuse negative magnetic area at the north. However, some electrons of them also precipitate towards the positive polarity, where absolute majority of protons precipitate. Most likely, this happens due to the presence of spatially-alternating electric fields.

\mrk{The main drawback of the combined MHD-particle approach using in this study is the lack of feedback: any electromagnetic field produced by propagating energetic particles is not included into the MHD simulations. However, relatively low acceleration efficiency in our model means that our results are within the validity range of the combined MHD-particle approach, since propagating energetic particles can be easily neutralised by the return current of ambient thermal electrons, thus maintaining charge neutrality on the MHD scale. Thus, 6$\times$10$^{36}$ electrons precipitating into the area of about 300~Mm$^2$ (3$\times$10$^{14}$~m$^2$, an area of the precipitation sites in the Figure~\ref{f-particles}) within about 200~s would produce the electron flux of about 10$^{20}$~m$^{-2}$~s$^{-1}$ (or the current density of about 0.16~A~m$^{-2}$). If these energetic electrons are not fully or partially neutralised by propagation ions, this flux would require the counter-flow of ambient electrons with the drift speed of about 10$^5$~m~s$^{-1}$ (assuming typical coronal density of 10$^{15}$~m$^{-2}$). Since this velocity is much smaller that the electron thermal speed in the corona ($\sim 10^7$~m~s$^{-1}$), thermal plasma would easily compensate the energetic electron precipitation without causing plasma instability.}

\begin{figure}[ht]
\centerline{\includegraphics[width=0.43\textwidth,clip=]{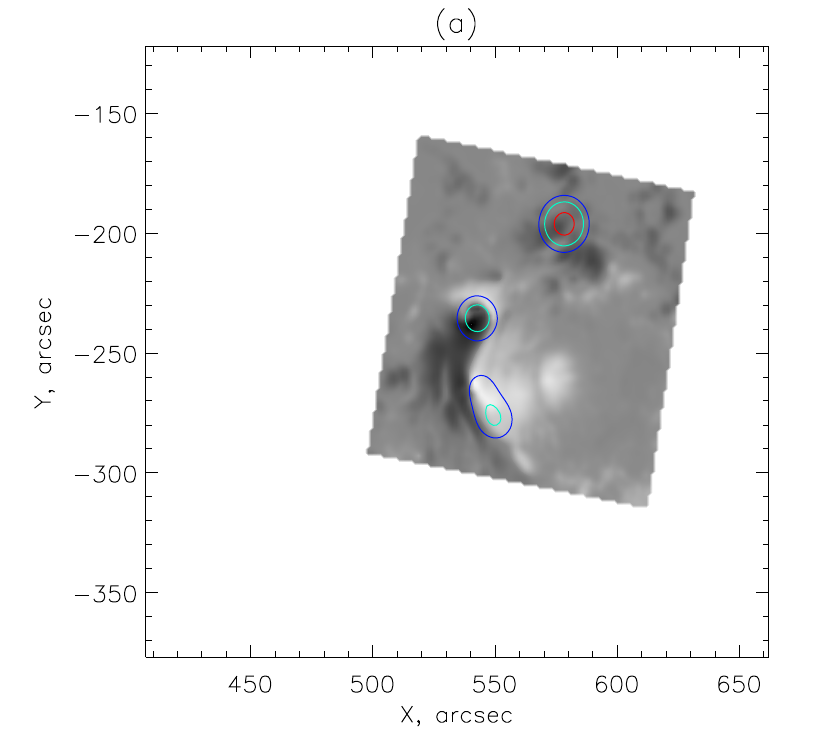}}
\centerline{\includegraphics[width=0.43\textwidth,clip=]{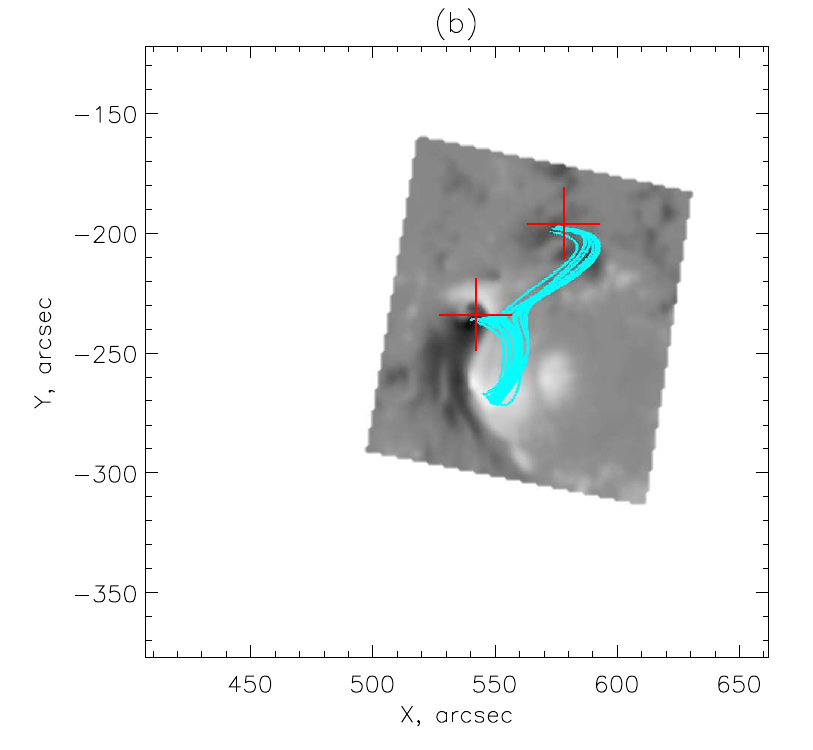}}
\centerline{\includegraphics[width=0.43\textwidth,clip=]{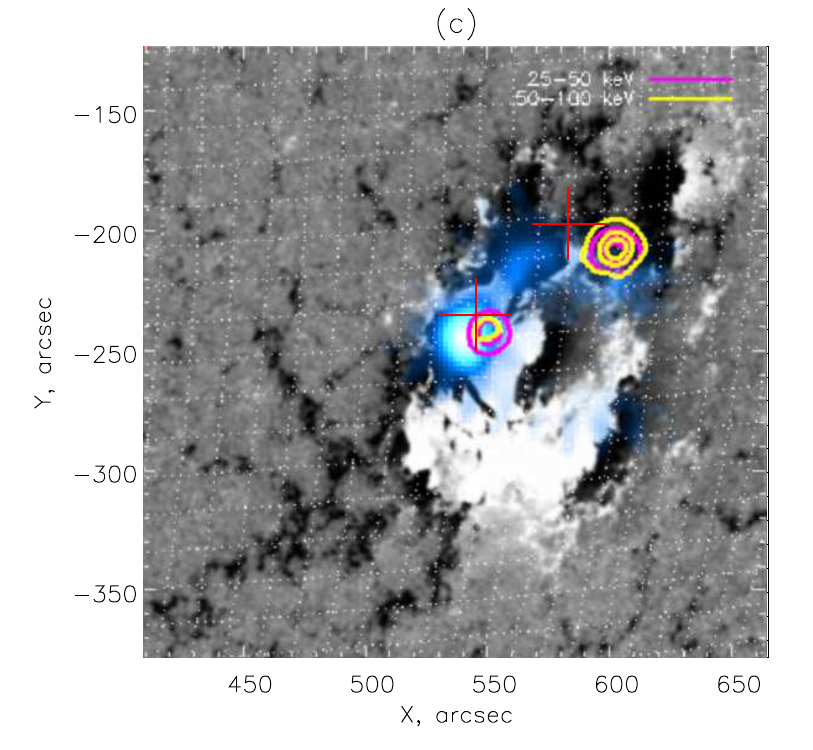}}
\caption{Panel (a) shows non-thermal bremsstrahlung intensity map at 8~keV at t=80~s, predicted by the model, over-plotted over the photospheric magnetic field map. Panel (b) shows the magnetic field line, along which the majority of energetic electrons propagate. Panel (c) shows contour plot of hard X-ray emission observed by RHESSI at 12:18~UT in the energy ranges 9--12~keV (blue area), 25--50~keV (pink line) and 50--100~keV (yellow line). The crosses in panels (b) and (c) show the centroids of two bright sources from the synthetic HXR map in panel (a).}
\label{f-rhessi}
\end{figure}

The typical travel time for energetic electrons is only about 1~s, which is much shorter than the evolution time-scale of the magnetic field in the MHD model (10--100~s). Therefore, the locations of bremsstrahlung X-ray emission sources, which coincide with the locations of the electron precipitation sites (right panels in Figure~\ref{f-particles}), are determined by momentary magnetic field configuration. Thus, there is a number of footpoint sources: one HXR source located over the negative sunspot, one source over the diffuse negative polarity in the NW part of the active region, a diffuse source over the positive sunspot, and a weak source over weak negative magnetic field in the NE corner.

This flare has been observed by the Reuven Ramaty High-Energy Solar Spectroscopic Imager (RHESSI) \citep{line02}, making it possible to compare our model predictions with observations. The observed HXR maps shown in Figure~\ref{f-rhessi} are similar, but not completely identical to the synthetic maps in Figure~\ref{f-particles}. Firstly, due to a limited number of test-particles in the present model, we can reliably calculate the non-thermal HXR sources only at energies up to about 8~keV, while RHESSI observes bremsstrahlung HXR with energes up to $\sim$100~keV. Secondly, the peak of this event was not observed by RHESSI for technical reasons. Still, we can compare the locations of the sources. It can be seen that two brightest sources predicted by our model are observed by RHESSI. The agreement between the locations is good: the difference is smaller than the size of the sources. This difference, in fact, may be partly due to the projection effect and non-zero height of the sources above the photosphere.

There is a number of possible reasons why the third diffuse bright source over the positive sunspot does not appear on the RHESSI map. Firstly, it may have a lower intensity at RHESSI energies. This, along with the diffuse shape of the source, may result in it being removed during the CLEAN process. Secondly, the RHESSI intensity map reflects a slightly later stage of the impulsive phase, and by this time the third source could disappear. Thirdly, it might be due to very strong polarity asymmetry of particle precipitation, as predicted by \citet{zhgo04}. This polarity asymmetry is clearly seen in our results, but there is some proton-electron mixing due to fragmented electric field structure. If the electric field in the real event has a solid structure with constant polarity, the asymmetry would be close to 100$\%$. Finally, our model depends on a number of assumptions, most importantly, on the functional form of anomalous resistivity, and these assumptions may affect the geometry of particle acceleration.

This flare has been recently studied by \citet{zhae20a} and \citet{zhae20b}, who calculated the locations of the helioseismic sources and estimated power required to produce observed helioseismic response. Three of their most powerful sources are found very close to some of the locations, where energetic particles precipitate in our model. Thus, one of the sources, produced by energy input of approximately $(6-8)\times 10^9$~J~m$^{-2}$~s$^{-1}$ is situated close to the location of $\cap$-shape of the neutral line (at $X=130$~Mm and $Y=130$~Mm in Figure~\ref{f-particles}, or in the centre of Figure~\ref{f-rhessi} a), where electrons precipitate. 

Two more powerful sources, found in the southern part of this region, appear to be close to the location, where some electrons and most protons precipitate in our model ($X=135$~Mm, $Y=130$~Mm and $X=140$~Mm, $Y=100$~Mm in Figure~\ref{f-particles}). The power required to create these sources is approximately $(8-12)\times10^9$~J~m$^{-2}$~s$^{-1}$ for each source. In other words, this heioseismic source required much higher energy input than other sources. This is an interesting finding, as it may indicate that energetic ions (or mixed ion-electron beams) are more effective in producing helioseismic response, compared to pure electron beams  \cite[e.g. see discussion in][]{gore05,zhzh07,mate15}.

\section{Summary}\label{s-summary}

In this study we have investigated particle acceleration and energetic particle transport in the evolving electromagnetic field of an actual solar flare observed in active region NOAA 12673 on the 6th of September, 2017. Analysis of the MHD model shows that the energy release and particle acceleration region in this flare is located relatively low, only about 15~Mm over the photosphere and has a cylinder-like shape with the cross-section diameter of about 5~Mm and length of around 20--25~Mm. 

The lifetime of the energy release region is around two minutes, and during this time it releases more than 10$^{25}$~J of energy, and accelerates around 6$\times$10$^{36}$ energetic electrons to the energies higher than 10~keV, carrying about 1.2$\times$10$^{23}$~J. Most of these energetic electrons precipitate to the photosphere, producing three bright footpoint sources of hard X-ray emission on the either side of the magnetic neutral line. The duration of the hard X-ray impulse (80--100~s) is in a good agreement with the typical duration of an impulsive phase in a solar flare.

Most importantly, comparison of our model predictions with the HXR imaging data from RHESSI shows that our model can relatively well predict the locations of chromospheric sources of non-thermal bremsstrahlung produced by precipitating energetic electrons.

Our model predictions are also in a good agreement with helioseismic observations of this event. Thus, three most powerful helioseismic sources detected by \citet{zhae20a} and \citet{zhae20b} are located very close to the particle precipitation sites in our test-particle simulations.

The present model is based on a number of assumptions, which may substantially affect particle kinetics. Obviously, these assumptions require further testing, by developing models of other events and comparing them with observational data. This research would be very timely now, since two new solar missions, Parker Solar Probe and Solar Orbiter, along with the LOFAR radio-telescope will provide unprecedented observations of solar energetic particles both in the solar corona and in the inner heliosphere.

\section*{ACKNOWLEDGMENTS}
This study has been supported by the International Collaborative Research Program of the Institute for Space-Earth Environmental Research (Japan). MG and PKB are funded by STFC (UK), grant ST/P000428/1. EPK acknowledges the financial support from the STFC Consolidated Grant ST/P000533/1. KK was supported by MEXT/JSPS KAKENHI Grant Numbers JP15H05814. Simulations have been performed partially using DiRAC Data Centric system at Durham University, operated by the Institute for Computational Cosmology on behalf of the STFC DiRAC HPC Facility.

\end{document}